\documentclass[10pt, conference]{IEEEtran}
\usepackage{graphicx}
\usepackage{subfigure}
\usepackage{multirow}
\usepackage{multicol}
\usepackage{algpseudocode}
\usepackage[]{algorithm2e}
\usepackage{enumerate}
\usepackage{flushend}
\usepackage[space]{cite}

\begin{document}

\title{N-Version Obfuscation: Impeding Software Tampering Replication with Program Diversity}

\author{\IEEEauthorblockN{
Hui Xu\IEEEauthorrefmark{1}\IEEEauthorrefmark{2},
Yangfan Zhou\IEEEauthorrefmark{2}\IEEEauthorrefmark{3},
Michael R. Lyu\IEEEauthorrefmark{1}
}
\IEEEauthorblockA{\IEEEauthorrefmark{1} Department of Computer Science,
The Chinese University of Hong Kong}
\IEEEauthorblockA{\IEEEauthorrefmark{2} MoE Key Laboratory of High Confidence Software Technologies (CUHK Sub-Lab)}
\IEEEauthorblockA{\IEEEauthorrefmark{3} Department of Computer Science, Fudan University}
}

\maketitle

\thispagestyle{plain}

\begin{abstract}

Tamper-resistance is a fundamental software security research area.  Many approaches have been proposed to thwart specific procedures of tampering, \textit{e.g.,} obfuscation and self-checksumming.  However, to our best knowledge, none of them can achieve theoretically tamper-resistance.  Our idea is to impede the replication of tampering via program diversification, and thus increasing the complexity to break the whole software system.  To this end, we propose to deliver same featured, but functionally nonequivalent software copies to different machines.  We formally define the problem as N-version obfuscation, and provide a viable means to solve the problem.  Our evaluation result shows that the time required for breaking a software system is linearly increased with the number of software versions, which is $O(n)$ complexity.  

\end{abstract}

\section{Introduction}

Once software has been released, it faces many security threats.  For example, attackers may crack its license protection mechanism to sell pirated software, or they may pack some payload (\textit{e.g.,} advertisement) into the original software for certain purposes.  Nowadays, the repacked app with malicious payload becomes even a major threat to mobile security \cite{zhou2012dissecting}.  Such attacks generally involve reverse engineering techniques to tamper the software.  To protect software from tampering, two major approaches have been proposed, \textit{i.e.,} obfuscation and self-checksumming.  On one hand, software can be protected with obfuscation approaches to deter attackers from locating the target code spot.  On the other hand, software can embed self-checksumming code to detect whether it has been tampered during execution.  Current tamper-resistant work mainly proposes such kinds of tricks to thwart specific tampering tools or approaches, such as using a loop with unsolved conjectures to thwart symbolic execution.  Nonetheless, once the trick is recognized, skillful attackers can design hand-crafted tools to launch an attack.  It seems safe to conjecture that software cannot achieve theoretically tamper-resistance without trusted hardware circuits \cite{chen2003oblivious}.

A general software tampering objective is to enable replicating the tampering on other machines.  Intuitively, we cannot guarantee a piece of software to be fully tamper-resistant, but we can fail the execution of tampered software on general machines, other than the attacker's.  Such an idea, known as program diversification \cite{forrest1997building}, is to prevent widespread attacks by making intrusions much harder to replicate.  If the attacker wish to run the tampered software on another machine, she has to work on it specifically.  In this way, we can cut off the contagion of tampering so as to control the scope of potential damage.

According to a recent survey \cite{larsen2015automatic}, existing software diversification approaches generally consider functionally equivalent programs, which can be effective against several kinds of attacks such as return oriented programming.  However, such approaches with the functionally equivalence constraint cannot meet our need to disable the replication of tampered software.  As a first attempt, we propose to deliver same featured, but functionally nonequivalent software copies to different machines.  A major challenge towards this goal is which part of a program can have such functionally nonequivalent diversities.  We formally define the problem as \textit{N-version obfuscation}, and provide a viable solution for the software of client-server architecture, \textit{i.e.,} by integrating a message authentication code (MAC) mechanism with functionally nonequivalent SHA1 algorithms \cite{sha1} to the original software, it can be resistant to tampering replication.  We further show that many software integrity protection problems can be reduced to our solution model.  It is worth noting that N-version obfuscation can be applied seamlessly to other existing tampering-resistant approaches, and hence equipping them with the replication-resistant property.  Our analysis result shows that the tampering complexity incurred by N-version obfuscation increases almost linearly with the number of functionally nonequivalent software versions.

The rest of this paper is organized as follows.  We first discuss the background in Section~\ref{section:motivation}.  We then demonstrate our approach in Section~\ref{section:approach}, and evaluate its effectiveness in  Section~\ref{section:evaluation}.  The related work is discussed in Section~\ref{section:literature}.  Finally, Section~\ref{section:conclusion} concludes this paper.

\section{Background} \label{section:motivation}
\subsection{Threat Model and Assumption} \label{section:threat}
Software delivered to end users is vulnerable to tampering.  Attackers may modify the original program execution logic for a specific purpose, and then replicate the tampering on other machines.  The modification can be achieved in two ways:

\textit{Software Repack}: Attackers can manipulate the software executables directly, for example, they may remove the original advertisement module embedded in a mobile app installation file, and replace it with another one beneficial to themselves.

\textit{Dynamic Injection}: Attackers may also dynamically inject a piece of code into the program process, so as to manipulate the loaded program during execution, \textit{e.g.,} using Linux \texttt{ptrace} tool.  Such an approach is widely adopted by virus and anti-virus software, which injects either inspection code to monitor the program execution or places a back door to control it.

In this work, we consider the hostile host model \cite{sander1998protecting}, which is widely adopted by software tamper-resistance work, such as \cite{horne2002dynamic,collberg2002watermarking}.  We assume that to launch such tampering attacks, attackers can use malicious host to analyze the software, and they can fully inspect the software execution step by step.

\subsection{Limitation of Anti-reverse Engineering} \label{section:threat}

Reverse engineering is a crucial technology for software tampering.  It involves a process that analyzes and manipulates a software based on its executables, \textit{e.g.,} in an executable and linkable format (ELF).  Anti-reversing techniques impedes such a process by adding tricks into the executables to fool the analyzer.  General reverse engineering on ELF files involves two phases: a disassembly phase, and an analyzing phase.  The disassembly phase decodes the ELF binaries to assembly code, which can be performed by some tools (\textit{e.g.,} IDA) automatically.  We can hardly impede the decoding because eventually the processor has to be able to decode and execute the file.  But on the other hand, many tricks have been proposed to obstruct the analyzing phase.  We discuss the major approaches and their limitations in what follows.

\begin{figure}[thb]
\centering
\subfigure[One way function]{
\label{fig:obf_owf}
\includegraphics[width=0.46\textwidth]{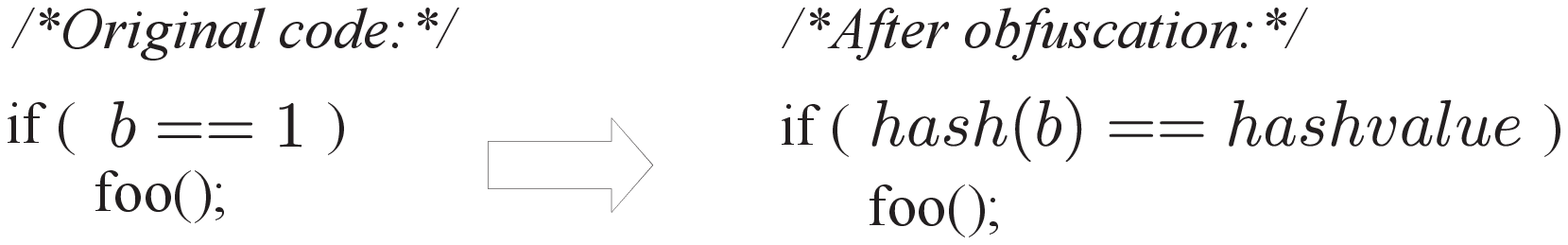}}
\subfigure[Opaque constant]{
\label{fig:obf_opq}
\includegraphics[width=0.17\textwidth]{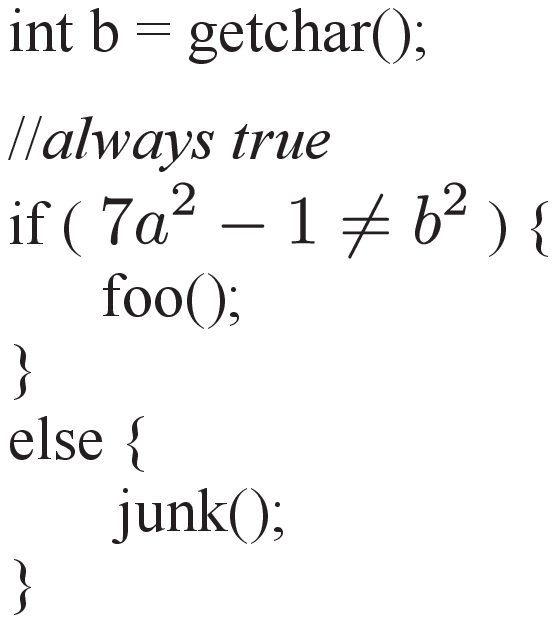}}\hfill
\subfigure[Collatz Conjecture]{
\label{fig:obf_uns}
\includegraphics[width=0.18\textwidth]{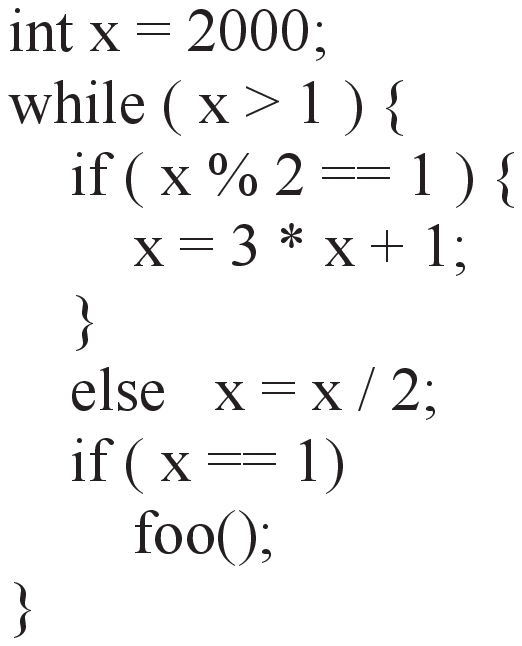}}
\caption{Demonstration of obfuscation approaches with different tricks. The original code for Fig.~\ref{fig:obf_opq} and Fig.~\ref{fig:obf_uns} is \textmd{\textit{foo();}}}
\end{figure}

There are two general ways to do reverse analysis, \textit{i.e.,} the offline approach and the online approach.  The offline approach does not execute the assembly code, but directly analyzes it using reverse engineering tools such as IDA \cite{ida}.  If a program has not been properly obfuscated, its control flow graph (CFG) can be easily derived, which shows the assembly code in blocks, and indicates their call relationships.  In this way, the complexity of analyzing the assembly code can be simplified.  CFG can provide great assistance for reverse engineers to grasp the meaning of the low-level assembly code which has little semantics.  To protect programs from being analyzed offline, a few obfuscation approaches have been proposed \cite{zhangprogram}.  The main idea of obfuscation is to add some junk code into the original program, while the original functionality of the program is still preserved.  Several methods have been proposed to achieve this.  For example, one can use opaque constant to add blocks of junk code that would never be executed.  Fig.~\ref{fig:obf_opq} shows such an example.  He may further create an NP-hard problem using a bunch of such code as discussed in \cite{ogiso2003software}.  Another method is to confuse the trigger condition of one code block with one-way function, so that the static analyzer cannot infer whether the code would be executed or is junk code.  Fig.~\ref{fig:obf_owf} shows an example of transferring an obvious condition into an opaque condition with a hash function.  Besides, unsolved conjectures have been proposed to confuse the exit criteria of loops, \textit{e.g.,} Fig.~\ref{fig:obf_uns} is an example of using Collatz Conjecture.  Such obfuscation techniques can increase great difficulties to general static analysis tools for analyzing the CFG and grasping the meaning of assembly codes.  However, all the obfuscation approaches have vulnerabilities.  For example, opaque constant is vulnerable to symbolic execution which implements a smart constraint solver, and unsolved conjectures are vulnerable to homemade tools which can recognize the patterns of conjectures.  

Even though many powerful offline analysis tools are available off-the-shelf, pure offline analysis still suffers hard limitations in detecting some anti-reversing protections, \textit{e.g.,} runtime code unpacking is widely used by malware to escape static analysis.  Therefore, adversaries may also execute the code to obtain execution instruction traces \cite{qiu2015identifying} or debug the code step by step to perform the analysis, which is known as online reverse-engineering \cite{eilam2011reversing}.  Such an analysis process generally would not be affected much by obfuscation \cite{yadegari2015generic}, and adversaries can leverage a set of system monitoring tools to monitor the execution outcome of a code block, which facilitates the reverse-engineering process.  Researchers have suggested to set traps with anti-debugger code to hinder debugging.  For example, one may simply check the debug register to detect if a debugger is present, or count the execution time of a code block to detect if it has been paused, and then penalize the debugger \cite{gagnon2007software,shields2010anti}.  Again, if the trick of anti-debugger code is recognized, adversaries can suppress the checking by patching the binaries, or simply switching to another debugger.  

When deriving enough understanding about the code, adversaries can manipulate the binaries by adding or deleting some code according to a specific purpose while preserving its ability of execution.  A possible way to detect such code patching is to use self-checksumming code.  The basic idea is to pre-calculate a value of relative address (\textit{i.e.,}, the checksum), and let the program fetch instructions during execution according to such a value.  If the checksum governed regions has been tampered, the instruction would not be correct and the program would likely to suspend \cite{wurster2005generic}.  Using overlapped self-checksumming code can further increase the strength of protection.  However, it can be defeated by carefully detecting and removing them \cite{qiu2014framework} or exploring the vulnerabilities \cite{wurster2005generic} of execution environment.

In other words, there is still no overwhelming anti-reverse engineering method, \textit{i.e.,} software can never be made fully resistant to tampering without hardware protection.

\section{Our Proposed Approach} \label{section:approach}
In this section, we introduce the N-version obfuscation problem first, and then discuss a possible solution with its application scenario in achieving tamper-resistance. 

\subsection{Problem Definition}
We formally define the N-version obfuscation problem as following: Given an algorithm $A$, how to automatically generate a large set of functionally nonequivalent algorithms $\{C_1,...C_n\}$, which are similar to $A$, and their parent algorithm $P$, so that they meet the following two properties:

\textit{Homo Property}: when performing on the same task, $P$ can output the same result as $C_i$, if the gene vector $\{g_1,...g_n\}$ of $C_i$ is known to $P$.  

\textit{Divergence Property:} when performing on the same task, $C_i$ and $C_j$ generally output different results. 

Suppose the software architecture is client-server mode, we can deploy the parent algorithm on the server side, and deliver a unique children algorithm to each client.  In this way, the software distributed to client can have such functionally nonequivalent diversities according to the divergence property, and the homo property enables the server to handle such diversities.

\subsection{N-Version Obfuscated SHA1}

In this section, we show a viable means to solve the N-version obfuscation problem with SHA1 algorithm.  Our approach leverages the iterations of calculations needed by SHA1 to generate functionally nonequivalent diversities.

The main loop of original SHA1 (Algorithm~\ref{alg:ori_sha1}) includes 80 rounds of iterations.  Each iteration takes one plaintext block ($w[i]$) into calculation.  For every twenty rounds, the calculation (the equation for generating $f$ and the value of $k$) switches to another one.  Even though there are some security considerations of choosing a specific calculation for each round, no evidence shows the program would suffer great security degradation if we switch them with each other.  Therefore, we can diversify the original SHA1 algorithm by choosing different sequences of equations for generating $f$ and values of $k$, which are the genes of individuals.  We can also design a parent algorithm which can receive the genes of a child, and process data input according to the setting of genes.  Algorithm~\ref{alg:alg_par} shows the such a parent algorithm we designed.  In Algorithm~\ref{alg:alg_par}, the pointer array of equations ($fp[80]$) for generating $f$, and the value array of $k$ for the 80 rounds of iterations are passed to the algorithm as the genes of a child.  It is clearly seen that, given the same input $w[80]$, the parent algorithm can compute the same result as a child when $fp[80]$ and $k[80]$ are properly set.

\begin{algorithm} [th]
\label{alg:ori_sha1}
\caption{The main loop of SHA1}
\small
\For {$i = 0; i < 80; i++$}	{
	\If {$0 \leq i \leq 19$}{
		$ f \gets (b$ AND $c)$ OR $(($NOT $b)$ AND $d) $\;
		$ k \gets$ 0X5A827999\;
	}
	\If {$20 \leq i \leq 39$}{
		$ f \gets b$ XOR $c$ XOR $d$\;
		$ k \gets$ 0X6ED9EBA1\;
	}
	\If {$40 \leq i \leq 59$}{
		$ f \gets (b$ AND $c) $ OR $(b $ AND $ d) $ OR $ (c $ AND $ d) $\;
		$ k \gets$ 0X8F1BBCDC\;
	}
	\If {$60 \leq i \leq 79$}{
		$ f \gets b $ XOR $ c $ XOR $ d $\;
		$ k \gets$ 0XCA62C1D6\;
	}
	$temp \gets (a$ LEFTROTATE $5) + f + e + k + w[i]$\;
	$e \gets d $\;
	$d \gets c $\;
	$c \gets b$ LEFTROTATE $30$ \;
	$b \gets a $\;
	$a \gets temp $\;
}
\end{algorithm}

\begin{algorithm} [th]
\label{alg:alg_par}
\caption{A parent algorithm for SHA1}
\SetKwProg{Fn}{Function}{}{}
\small
\KwData {$fp[80], k[80], w[80]$}
\For {$i = 0; i < 80; i++$}{
	Call $fp[i]$\;  \tcp*[h]{Pointer to F0, F1, F2 or F3}
	F\_TAIL$(k[i], w[i])$\;
}

\Fn{F0()}{
	$ f \gets (b$ AND $c)$ OR $(($NOT $b)$ AND $d) $\;
}
\Fn{F1()}{
	$ f \gets b$ XOR $c$ XOR $d$\;
}
\Fn{F2()}{
	$ f \gets (b$ AND $c) $ OR $(b $ AND $ d) $ OR $ (c $ AND $ d) $\;
}
\Fn{F3()}{
	$ f \gets b $ XOR $ c $ XOR $ d $\;
}
\Fn{F\_TAIL($k, w$)}{
	$temp \gets (a$ LEFTROTATE $5) + f + e + k + w$\;
	$e \gets d $\;
	$d \gets c $\;
	$c \gets b$ LEFTROTATE $30 $\;
	$b \gets a $\;
	$a \gets temp $\;
}
\end{algorithm}

\subsection{Implementation}
We automate the process of generating N-version SHA1 algorithms based on Low Level Virtual Machine (LLVM) \cite{llvm}, which is a widely used opensource compiler that supports extensions.  LLVM first represents the source code with Abstract Syntax Tree (AST), and then transfers it into intermediate code (IR), which would finally be compiled into executables according to a specific platform.  LLVM supports \texttt{plugin} and \texttt{libtooling}, which can manipulate the source code of a target AST branch during the compilation process.  We hence customize a \texttt{libtooling} tool that can automatically generate the N-version obfuscation algorithms.  

According to Algorithm \ref{alg:alg_par}, each gene (either $fp[i]$ or $k[i]$) has four possibilities, so we use two bits to represent a gene.  In each compilation, we first randomly generate two 160 length bit sequences: one as the chromosome for the equation function pointer (\textit{i.e.,} $fp[80]$) and the other as the chromosome for the value option of $k$ (\textit{i.e.,} $k[80]$).  We then replace the corresponding AST branches with hardcoded equation function pointers and settings of $k$.  

It is worth noting that the N-version obfuscation approach itself does not provide any resistance to reverse engineering.  However, our approach can be seamlessly integrated with other aiti-reverse engineering protections, for example, the obfuscation approaches proposed in \cite{ogiso2003software}, which composes NP-hard problems with function pointers and opaque constants.  

\subsection{Application Discussion} \label{section:application}

\begin{figure}[t]
\centering
\includegraphics[width=0.46\textwidth]{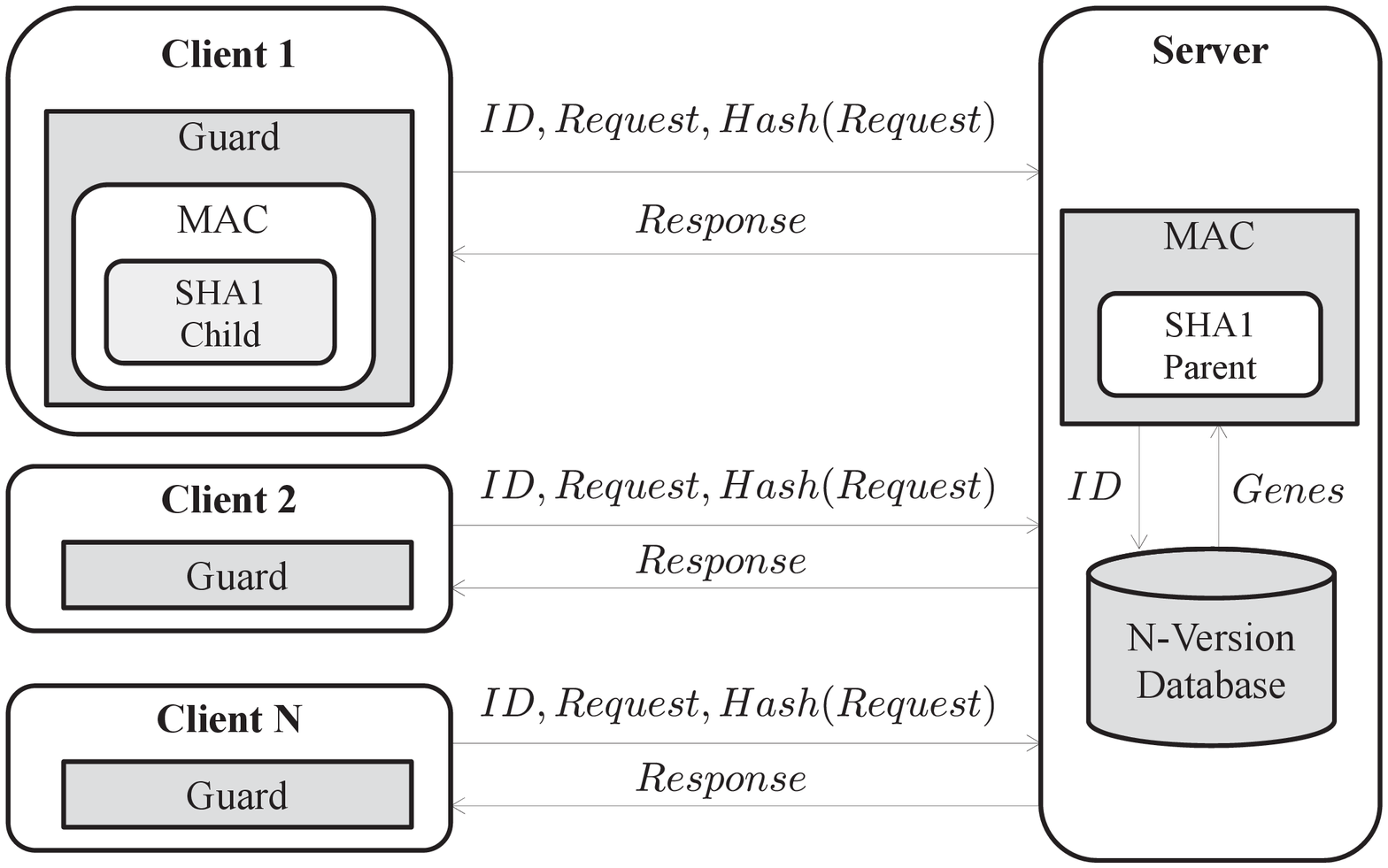}
\caption{Application of N-version obfuscated program in tamper resistance}
\label{fig:application}
\end{figure}

Many client-server software can adopt the N-version obfuscation idea by implementing a MAC mechanism with the obfuscated algorithm (\textit{e.g.,} SHA1).  MAC is a popular mechanism adopted by client-server computing architecture to check the integrity and authenticity of messages.  When a client sends a request to the server, it calculates the MAC of the request and appends it to the original request.  The server validates the MAC first and then processes the request.  Since a hash function is one major component of a MAC algorithm, the N-version obfuscated SHA1 algorithms can be adopted.  Fig.~\ref{fig:application} illustrates such a mechanism.  

In Fig.~\ref{fig:application}, each client is embedded with a unique SHA1-based MAC calculation algorithm.  To successfully perform a request to the server, it has to send the identification (such as machine serial number or user id), the request, and the MAC together to the server.  The server queries the genes of a client from its local N-version database according to the identification of the client, and then verifies the MAC.  The distribution of such diverse programs can be achieved by implementing the MAC in mobile code (\textit{i.e.,} dynamic library), and delivering it by the server upon request.  In other words, the client software can be launched without the library at the first time and then requests the server for the library.  The server randomly chooses a library from a pool of pre-compiled libraries and delivers it to the client; in the meanwhile, it records the mapping between the genes of the client and its unique identification in the N-version obfuscation database.  In this way, the server can verify the MAC generated by each client according to the homo property.

\begin{figure}[t]
\centering
\includegraphics[width=0.48 \textwidth]{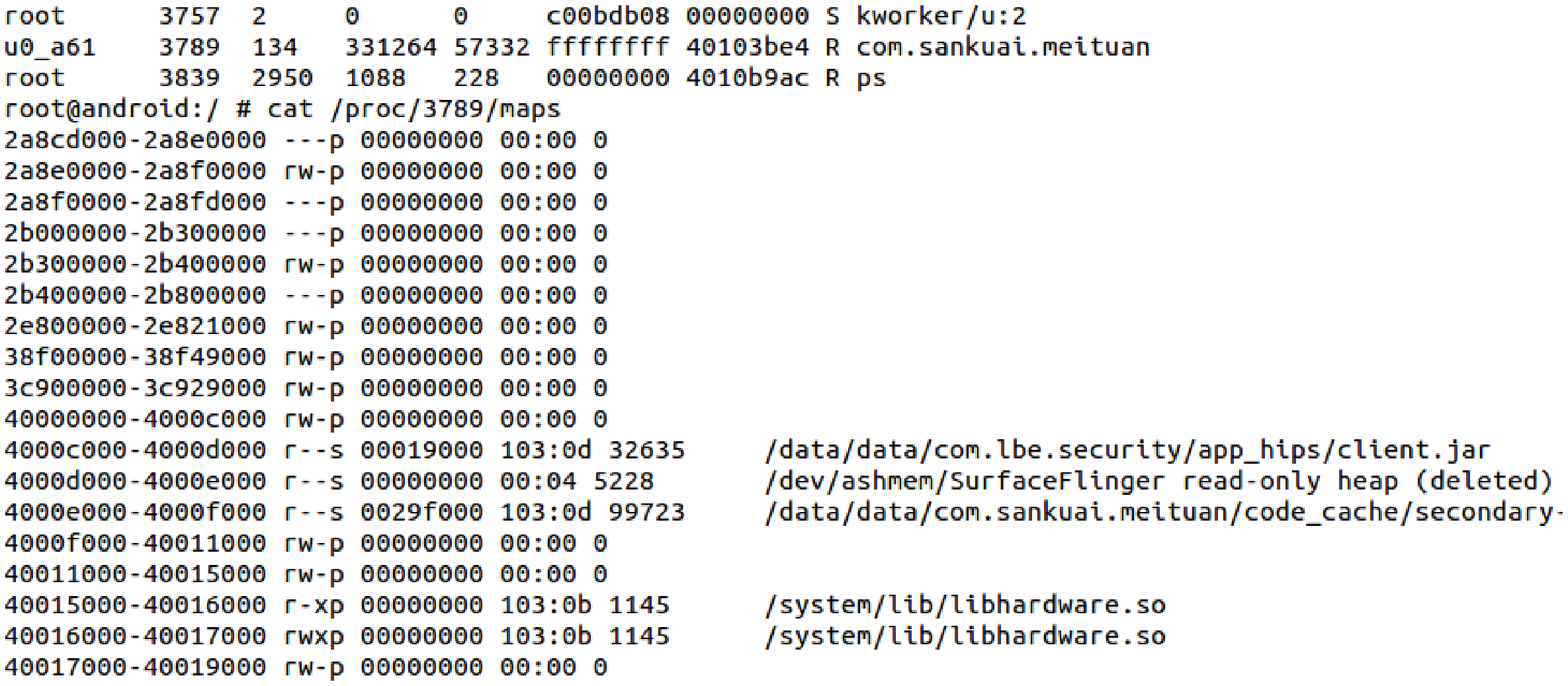}
\caption{An exemplary Android app (\texttt{com.sankuai.meituan}), which has been injected with an LBE library (\texttt{com.lbe.../client.jar}).  By checking the maps file of the app process (pid:3789), we can detect the tampering.}
\label{fig:lbe}
\end{figure}

We hence provide tamper-resistant capability for the client software based on the N-version obfuscation library, which is resistant to replication according to the divergence property.  To this end, a viable means is to implement an integrity checking function align with the MAC in the library, so that the library can serve as a security guard for the software.  By interleaving the code of the integrity checking function with the MAC algorithm, the integrity checking can be triggered when calculating a MAC.  Algorithm~\ref{alg:int_chk} shows such an exemplary integrity checking function for the apps of Android operating system.  The function navigates the maps file of the app process itself, which records the program segments and their address in the memory.  It then compares the record with a previously defined standard dictionary by developers.  If there are abnormal segments in the maps, \textit{i.e.,} the integrity has been violated, a responsive mechanism can be triggered.  Such an approach is effective in detecting either software repacking or dynamic injection attack as we have discussed in the adversary model.  We show an exemplary App (Fig.~\ref{fig:lbe}), which is tampered by LBE (a commercial security software for Android \cite{lbe}).  Algorithm~\ref{alg:int_chk} can detect such a tampering by finding that \texttt{com.lbe.../client.jar} is an abnormal segment.

\begin{algorithm} [t]
\caption{An exemplary integrity checking function}
\label{alg:int_chk}
\small
\KwData {$dict<segment>$}  \tcp*[h]{A list of predefined segment with name and size}\\
\SetKwProg{Fn}{Function}{}{}
\Fn(){IntegrityChk()}{}
	$pid \gets$ getpid()\;
	$file \gets$ open (/proc/pid/maps)\;
	\While {$ line \gets$ readline($file$) != EOF} {		
		$segName \gets$ GetSegName($line$)\;
		$segSize \gets$ GetSegSize($line$)\;
		\eIf {$!dict$.contains($segNmae$)}{
			Reaction()\;
		}
		{
			\If{$dict$.getsize($segNmae$)!=$segSize$}{
				Reaction()\;
			}
		}
	}
\end{algorithm}

If an attacker has successfully tampered one copy of the guard (\textit{e.g.,} removing the integrity checking function) and replicated it on other machines, the server can detect the replication because of an incorrect MAC, \textit{i.e.,} inconsistent mapping between the identification and genes.  We may further implement a reaction mechanism to renew the guard or crash the client software directly.  

A question to ask is why we do not simply use different keys to compose diversities?  For example, we may use a keyed-hash message authentication code (HMAC) algorithm and hardcode a unique symmetric key into each client library.  Note that such an approach is also effective, but it is more vulnerable than ours, because hiding a key (\textit{i.e.,} whitebox cryptography) is more difficult than hiding the program logic \cite{chow2003white}.

\section{Evaluation} \label{section:evaluation}
The goal of our work is to impede tampering replication by creating diverse software instances, and thus to increase the tampering complexity to the software system.  In this section, we evaluate the complexity increased by N-version obfuscation for tampering multiple software clients.

Suppose a program has adopted the protection mechanism discussed in Section \ref{section:application}.  A decent attacker wishes to manipulate the program binaries for a specific purpose, including adding, deleting and modifying.  According to the adversary analysis,  the guard cannot be simply removed or disabled from the app, because the MAC mechanism rested in the guard needs to be executed.  However, In a hostile host environment, the software can be fully inspected.  Through careful analysis, the attacker can identify that the protection lies in the integrity checking function, \textit{i.e.,} Algorithm~\ref{alg:int_chk}.  She can disable the checking by carefully removing the function or suppressing the reaction.  If there is no N-version obfuscation protection, the attacker can deliver the tampered software to other users.  The replication of tampered software can be executed normally on other machines, and the whole software system is broken.  With N-version obfuscation protection, if the attacker simply repacks the app with tampered guard, the message verification would fail, and the program cannot function normally.  To break the whole software system, the attacker need to analyze and tamper the guard of each machine specifically.  

In order to tamper the software on another machine, the attacker has to obtain the guard on that machine, develop a tampered guard, and then substitute the original one.  If the guard is protected with interleaved self-checksumming code \cite{chang2002protecting}, a successful tampering requires removing all the self-checksumming code at the same time, of which the chance is very low without sophisticated analysis.  Existing approaches on identifying such code generally require using dynamic analysis and taint analysis together \cite{qiu2014framework}.  Empirically, the time required to tamper each guard is not negligible.

Let $c_0$ denote the time needed for analyzing one software copy and tampering it on the attacker's own hostile host.  The time complexity is $O(1)$, which is equal to tampering one software copy without N-version obfuscation.  Let $c_1$ denote the time needed to obtain the guard on another machine, so as to replace it, and $c_2$ denote the time needed to tamper it.  If the attacker wishes to tamper the software on $n$ machines, the total time can be estimated as $c_0 + n * (c_1 + c_2)$.  Because of the interleaved self-checksumming code, $c_2$ should not be negligible, hence the complexity equals to $O(n)$. 

Another possible tampering approach is to build an algorithm similar to the parent algorithm, which can calculate the hash value according to the genes of a specific child.  To this end, the attacker has to recognize the gene setting of each guard.  She may compare the difference between two implementations, and locate the genes.  If the attacker has derived enough knowledge about our N-version obfuscation theory and implementation, such kind of attack is theoretically possible.  However, if the guard is obfuscated (\textit{e.g.,} with \cite{ogiso2003software}), it would be very difficult.  To our best knowledge, existing work on breaking such obfuscated programs either require symbolic execution with sophisticated constraint solvers or complicated taint analysis \cite{yadegari2015generic}, which is computational intensive and time-consuming.  Let $c_3$ denote the time needed to extract the genes of a guard.  The time needed to tamper the software system can be estimated to $c_0 + n * (c_1 + c_3)$.  Because $c_3$ is not negligible, the complexity still equals to $O(n)$.  Note that $n$ can be made arbitrally large as the obfuscation task can be fully automated.

\section{Related Work} \label{section:literature}

In this section, we first review the important work in reverse engineering field, and then discuss the work of applying automated program diversity to improve program security.

\subsection{Reverse Engineering}
Software protection is a research problem since decades ago.  The proposed solutions are generally two-fold: the hardware circuit assisted solutions which provide better security assurance, or the pure software solutions which have better adaptability to general hardware \cite{chen2003oblivious}.  For our research problem, hardware circuit assisted solutions are not applicable because of their requirement on specific hardware, so we mainly discuss the pure software solutions.

Literatures on software protection with anti-reverse engineering approaches have different purposes.  While some researchers look for protections against piracy \cite{linn2003obfuscation,ogiso2003software} and intrusion \cite{chang2002protecting}, others investigate on impeding malware against detection \cite{moser2007limits,o2011obfuscation,sharif2008impeding}.  However, they share a set of common protection techniques with only slightly difference.  Obfuscation is a basic software protection approach.  It can complicate the binaries, and increases the difficulty of the reverse engineering.  Ogiso \textit{et al.} propose to obfuscate the code by constructing a NP-Hard complexity problem, that requires to determine the real function pointer from an array of pointers \cite{ogiso2003software}.  However, such an approach is vulnerable to symbolic execution with constraint solvers.  To thwart symbolic executions, Sharif \textit{et al.} notice that some code blocks can be concealed by setting a trigger condition with an one-way function, so that the constraint solver cannot solve \cite{sharif2008impeding}.  Wang \textit{et al.} propose another obfuscation technique to combat the symbolic execution by exploring the general limitation of symbolic execution tools in analyzing loops.  Their idea is to use unsolved conjectures \cite{wang2011linear} to confuse the termination condition of loops.  The approach is vulnerable when the tricks of unsolved conjectures are recognized.  Other than setting tricks on the source code, Linn \textit{et al.} propose to obfuscate the binaries directly by inserting some error bits, which can be automatically corrected during execution by the CPU but not by current disassembly tools \cite{linn2003obfuscation}.  The security of such a protection is very limited and vulnerable to dynamic analysis, \textit{i.e.,} the actual instruction trace can be easily obtained once the software is being executed.  To deter dynamic analysis with debuggers, Oishi \textit{et al.} propose to use some camouflaged anti-debuggers, which, however, is not effective for homemade debuggers.  On protecting software from tampering, another general popular approach is to detect the unauthorized modifications during runtime by employing a self-checksumming mechanism \cite{horne2002dynamic, chang2002protecting}.  The self-checksumming mechanism uses redundantly overlapped checksum testers inside the program to verify its integrity.  On the other side, several investigations focus on defeating the protections \cite{wurster2005generic,moser2007limits,qiu2014framework,yadegari2015generic}.  Wurster \textit{et al.} propose to defeat the self-checksumming approach with a duplicated memory attack, and examine its effectiveness on several popular cpu types \cite{wurster2005generic}.  Qiu \textit{et al.} propose to identify the self-checksumming code using taint analysis approaches\cite{qiu2014framework}.  Yadegari \textit{et al.} find a more general way of deobfuscating an obfuscated algorithm \cite {yadegari2015generic}.  Our work is different from all the existing work in that we focus on impeding the replication of software tampering.

\subsection{Automated Software Diversification}

The idea of software diversity is initially proposed for software reliability engineering \cite{lyu1996handbook}.  Cohen in \cite{cohen1993operating} firstly proposes to create functionally equivalent programs to enhance software security.  Forrest \textit{et al.} in \cite{forrest1997building} also states that the beneficial effects of diversity in computing systems have been overlooked, and introducing diversities into computer systems can make them more robust to replicated attacks.  They propose several possible ways to create such diversities with respect to the program behavior, including adding nonfunctional code, refectoring code, or diversifying the memory layout.  Crane \textit{et al.} in \cite{crane2015readactor} build upon fine-grained code diversification to prevent code-reuse attacks.  They adopt function permutation \cite{kil2006address}, register allocation randomization, and callee-saved register save slot reordering \cite{pappas2012smashing} in the diversification process.  However, to our best knowledge, all these work do not consider to automatically generate functionally nonequivalent programs to improve tamper-resistance.

\section{Conclusion} \label{section:conclusion}
This work has investigated software tamper-resistant issues with N-version obfuscation.  We have formally defined the N-version obfuscation problem and provided a viable solution with SHA1 algorithm.  We have further discussed the application of such an N-version obfuscation idea in the client-server software architecture.  By introducing such functionally nonequivalent diversities, our approach is effective to impede the replication of tampering.  The evaluation result shows that the complexity to tamper the software system is linearly increased with the number of software versions, which can be automatically generated with trivial cost.

\bibliographystyle{abbrv}
\bibliography{trap}

\end{document}